\documentclass[%
 preprint,
superscriptaddress,
 amsmath,amssymb,
 aps,
]{revtex4-2}

\usepackage{graphicx}% Include figure files
\usepackage{dcolumn}% Align table columns on decimal point
\usepackage{bm}% bold math
\usepackage{mhchem}
\begin{document}

%\preprint{AIP/123-QED}

\title{First-principles analysis of the optical properties of lead halide perovskite solution precursors}
% Force line breaks with \\

\author{Giovanni Procida}
\affiliation{Carl von Ossietzky Universit\"at Oldenburg, Institute of Physics, 26129 Oldenburg, Germany}
\author{Richard Schier}
\affiliation{Humboldt-Universit\"at zu Berlin, Physics Department and IRIS Adlershof, 12489 Berlin, Germany}
\author{Ana M. Valencia}
\affiliation{Humboldt-Universit\"at zu Berlin, Physics Department and IRIS Adlershof, 12489 Berlin, Germany}
\affiliation{Carl von Ossietzky Universit\"at Oldenburg, Institute of Physics, 26129 Oldenburg, Germany}
\author{Caterina Cocchi}
\affiliation{Humboldt-Universit\"at zu Berlin, Physics Department and IRIS Adlershof, 12489 Berlin, Germany}
\affiliation{Carl von Ossietzky Universit\"at Oldenburg, Institute of Physics, 26129 Oldenburg, Germany}
\email{caterina.cocchi@uni-oldenburg.de}


\begin{abstract}
Lead halide perovskites (LHPs) are promising materials for opto-electronics and photovoltaics, thanks to favorable characteristics and low manufacturing costs enabled by solution processing. In light of this, it is crucial to assess the impact of solvent-solute interactions on the electronic and optical properties of LHPs and of their solution precursors. In a first-principles work based on time-dependent density-functional theory coupled with the polarizable continuum model, we investigate the electronic and optical properties of a set of charge-neutral compounds with chemical formula, PbX$_2$(Sol)$_4$, where X=Cl, Br, and I, and Sol are the six common solvents. We find that single-particle energies and optical gaps depend on the halogen species as well as on the solvent molecules, which also affect the energy and the spatial distribution of the molecular orbitals, thereby impacting on the excitations. We clarify that dark states at the absorption onset are promoted by electron-withdrawing solvents, and we show the correlation between oscillator strength and HOMO$\rightarrow$LUMO contribution to the excitations. Our results provide microscopic insight into the electronic and optical properties of LHP solution precursors, complementing ongoing experimental research on these systems and on their evolution to photovoltaic thin films. 
\end{abstract}

\maketitle
\newpage
%%%MAIN TEXT%%%%

\section{Introduction}\label{intro}

The power conversion efficiency of solar cells based on lead halide perovskites (LHP) has increased within a decade up to the current record of 29\%~\cite{alas+20sci}.
Such an impressive performance is related to the intrinsic characteristics of these materials, including their large optical absorption coefficients, their high mobility, their long balanced carrier diffusion length, as well as their low exciton binding energies~\cite{stra+13sci,dinn+14natcom,shal+16acsami,xiao-yen17aem,yang+17jpcl,umar+18jpcl,lim+19ees}.
These extraordinary opto-electronic properties are supported by the ability of LHPs to be synthesized by solution processing~\cite{li+15jmca,yang+15aem,hami+17acsel,mcdo+18am,jung+19csr,soto+20aplm,kim+20jacs,zhou+21aom}, which enables reduced production costs, high chemical tunability, and straightforward thin-film deposition~\cite{jeon+14natm,wu+14ees,zhan+16jmca,daen+18jpcl,huan+19jmca}. 
However, a critical aspect of this preparation method is that the nature of the solvents and their coupling with the solute may impact on the characteristics of the final products~\cite{vayn20aem}.
A number of recent studies addressing the stability of LHP solution precursors have confirmed that solute-solvent interactions influence the electronic and optical characteristics of the resulting thin films~\cite{rahi+16chpch,yoon+16jpcl,stev+17cm,shar+17cm,radi+19acsaem,radi+20jpcl,sore+21jmca}.
Additional works investigating intermediate phases have contributed to elucidate the key mechanisms ruling the formation of these materials and their evolution to the crystalline form~\cite{mans+15jpcc,guo+15jacs,cao+16jacs,petr+17jpcc,fole+17jmca,li+18cgd,fate+18cm,li+19small,shar+20ma,vasq+21jec,kais+21ma}.

Besides these important findings, a number of questions remain open regarding the electronic properties of LHP solution precursors and the microscopic understanding of their optical response.
While it is well established that actual samples contain various lead halide species with different charge and coordination numbers~\cite{rahi+16chpch,radi+19acsaem,shar+20chpch,shar+20ma}, including seed structures of iodoplumbate chains~\cite{vale+21jpcl},
the identification of their spectral signatures in solution as well as in subsequent (pseudo)crystalline stages of formation is very challenging, due to the manifold of coexisting structures and their non-trivial mutual interactions~\cite{yu+14dt,mans+15jpcc,yoon+16jpcl,shar+20chpch,vasq+21jec}.
Given the intrinsic quantum-mechanical nature of the involved processes, experimental characterization alone is insufficient to clarify this intricate scenario.

Atomistic simulations have recently provided important insight into the formation of LHP solution precursors and their stability in different environments~\cite{radi+19acsaem,radi+20jpcl,schi+21pssb} .
However, there is still a lack of knowledge (i) on the electronic coupling between lead and the various halogen atoms; (ii) on the way these interactions affect the electronic structure of the materials; (iii) on the role of solvent species in altering these properties and in particular the optical response of LHP solution precursors.
Addressing these points thoroughly is essential to complement experiments in the rationalization of the observed trends for excitation energies and oscillator strengths in these complex systems, where optical absorption spectra are usually not prone to straightforward interpretation~\cite{shar+20ma,vale+21jpcl}.
Such knowledge is needed to optimize the preparation recipes of LHP solution precursors as well as to predict and understand their evolution process towards efficient photovoltaic thin films.

\begin{figure}[h]
	\centering
	\includegraphics[width=0.65\textwidth]{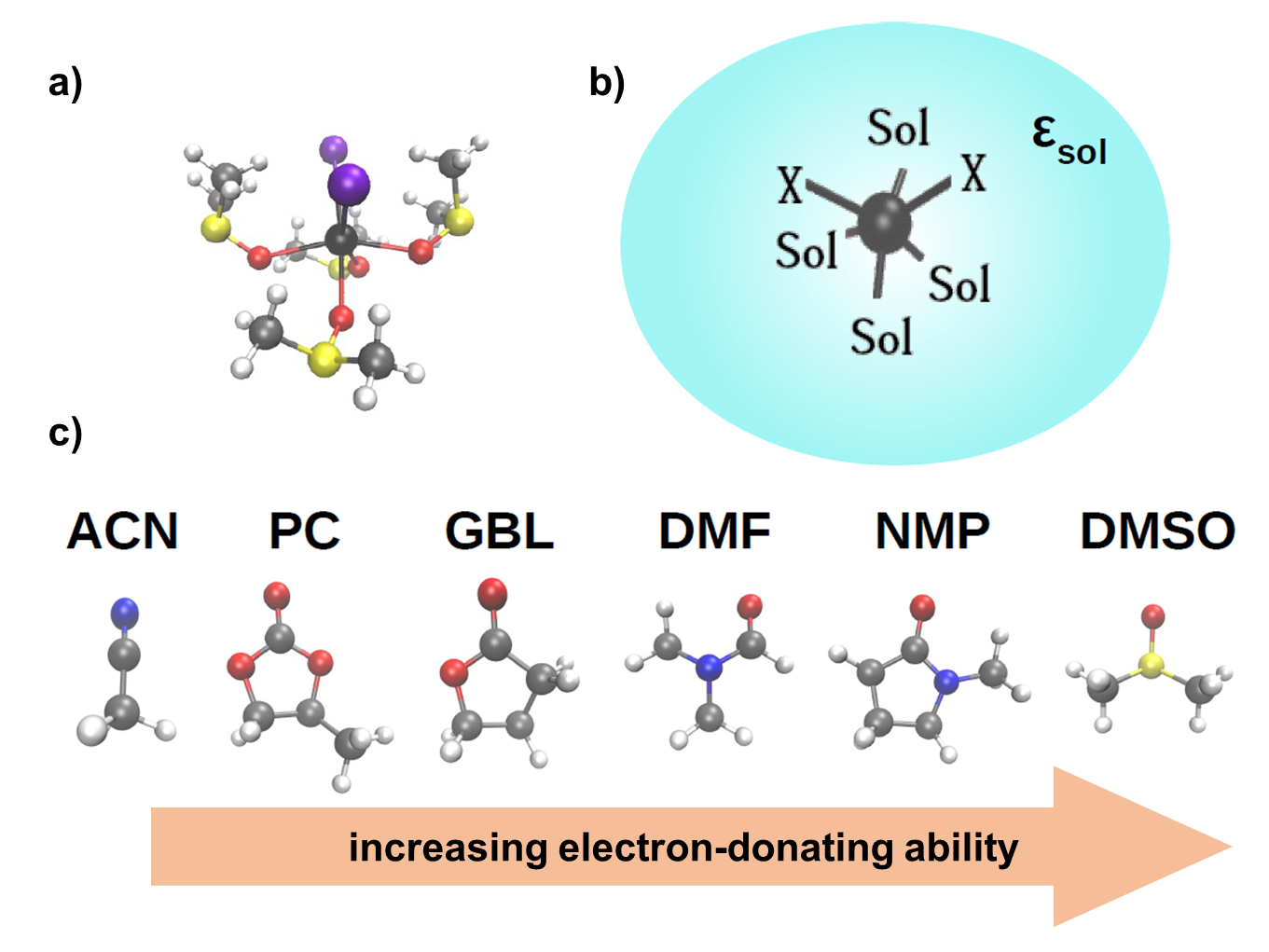} 
	\caption{a) Ball-and-stick representation of PbI$_2$DMSO$_4$ as exemplary compound investigated in this work; b) Pictorial sketch of PbX$_2$(Sol)$_4$ complexes in a solvent cavity with dielectric function $\varepsilon_{sol}$; c) Ball-and-stick representations of the solvent molecules considered in this work ordered from left to right according to their increasing electron-donating ability. C atoms are depicted in grey, H in white, O in red, N in blue, S, in yellow, I in purple, and Pb in dark grey.}
	\label{fig:solvents}
\end{figure}

In first-principles work based on hybrid time-dependent density functional theory (TDDFT) coupled to the polarizable continuum model (PCM), we investigate the electronic and optical properties of a representative set of LHP solution precursors including the charge-neutral compounds with chemical formula PbX$_2$(Sol)$_4$ (see Fig.~\ref{fig:solvents}a and Fig.~S1 in the Supporting Information), where X = Cl, Br, I, and Sol stands for the six solvent molecules C$_2$H$_3$N (ACN), C$_3$H$_7$NO (DMF), C$_2$H$_6$OS (DMSO), C$_4$H$_6$O$_2$ (GBL), C$_5$H$_9$NO (NMP). 
With the adopted \textit{ab initio} formalism, where both quantum-mechanical as well as electrostatic solvent-solute interactions are accounted for,  we are able to clarify how the considered solvent molecules chemically bound to the lead halide backbone affect the energies and the spatial distribution of the single-particle electronic states in the complexes.
With this insight, we compute and analyze the optical absorption spectra, focusing specifically on the lowest-energy excitations, which, depending on the electron-donating activity of the solvent and on their composition in terms of orbital transitions, can be optically active or almost dark, thereby influencing the photo-response of the LHP solution precursor.
Our findings indicate that both, halogen species and solvent molecules impact on the electronic structure and the optical response of the solution complexes.

%

%%%%%%%%%%%%%%%%%%%%%%%%%%%%%%%%%%%%%%%%%

\section{Methodology}

\subsection{Theoretical Background}
The results of this work are obtained from TDDFT~\cite{rung-gros84prl} employing a range-separated hybrid functional (see Section~\ref{sec:comput}).
The linear-response Casida scheme~\cite{casi-huix12arpc} is adopted in combination with the PCM~\cite{toma+05cr} to compute optical absorption spectra accounting for the dielectric screening exerted by the considered solvent molecules onto the PbX$_2$ core. 
In the PCM framework, the solute is immersed in a homogeneous cavity in which the solvent is modelled implicitly by its dielectric constant, $\epsilon_{Sol}$ (see Fig.~\ref{fig:solvents}b). 

The starting point of these calculations is the solution of the time-independent Kohn-Sham (KS) equations~\cite{kohn-sham65pr}, which, coupled to the PCM and in atomic units, read:
\begin{equation}
	\biggl[\frac{1}{2}\nabla^{2}+ V_{Hxc}[\rho]({\mathbf r})+V_{ext}({\mathbf r})+V_{PCM}[\rho]({\mathbf r}) \biggr]\phi_{i}^{KS}({\mathbf r}) = \epsilon_{i}\phi_{i}^{KS}({\mathbf r}),
	\label{eq:ks}
\end{equation}
where $\rho(\mathbf{r})=\sum_{i=1}^{N}|\phi^{KS}_i(r)|^{2}$
is the electronic density generated by the $N$ electrons in the system, and $\phi^{KS}_i$({\textbf{r}}) and $\epsilon_i$ the eigenfunction and eigenenergy of $i$-th KS state, respectively.
In Eq.~\eqref{eq:ks}, three potential terms appear: V$_{Hxc}$[$\rho$]({\bfseries r}) is the combined
Hartree exchange$-$correlation (Hxc) potential, which takes
into account electron$-$electron interactions and is herein approximated by the range-separated hybrid functional CAM-B3LYP~\cite{yana+04cpl}; V$_{ext}$(\textbf{r}) is the external potential embedding the Coulomb attraction from the nuclei to the electrons; V$_{PCM}$[$\rho$](\textbf{r}) is the effective potential describing the electrostatic interactions between solute and solvent.

To access excited-state properties, Eq.~\eqref{eq:ks}  has to be generalized to the time domain~\cite{rung-gros84prl}.
In the adopted linear-response formalism~\cite{casi-huix12arpc}, the problem is mapped into the eigenvalue equation, written below in compact matrix form as
\begin{equation}
	\begin{pmatrix}
		A & B \\
		B^{*} & A^{*}  
	\end{pmatrix}
	\begin{pmatrix}
		X  \\
		Y  
	\end{pmatrix}=\omega \begin{pmatrix}
		-1 & 0 \\
		0 & 1  
	\end{pmatrix}\begin{pmatrix}
		X  \\
		Y  
	\end{pmatrix},
\end{equation}
where the eigenvalues $\omega_i$ are the excitation energies, and the eigenvectors, $X^{i}_{ak}$ and $Y^{i}_{ak}$, are the resonant and antiresonant contributions, respectively, to the \textit{i}-th excitation from the single-particle transitions between occupied (\textit{k}) and unoccupied (\textit{a}) states.

TDDFT coupled to range-separated hybrid functionals (see Section~\ref{sec:comput} below) is known to be highly reliable in the calculation of the electronic and optical properties of complex molecules~\cite{wong-hsie10jctc,refa+11prb,bhan-duni19jctc}.
The possibility to be combined with PCM in order to account for solvation effects makes it the method of choice to rationalize the opto-electronic properties of complex systems in solution (see, \textit{e.g.}, our recent experimental/computational study on doped organic semiconductors, Ref.~\cite{arvi+20jpcb}).
For comparison, we report in the Supporting Information (SI) and discuss in Section~\ref{ssec:opt} the results obtained \textit{in vacuo} for PbI$_2$(Sol)$_4$ adopting many-body perturbation theory, the state-of-the-art approach to compute electronic and optical excitations in gas-phase molecules and in their crystalline counterparts~\cite{cocc-drax15prb}.

\subsection{Computational Details}
\label{sec:comput}
All calculations are performed with the code Gaussian 16~\cite{g16}.
Van der Waals interactions are accounted
for through the semi-empirical Grimme D3 dispersion scheme~\cite{grim+10jcp}.
In the structural optimizations, the cc-pVDZ basis set is employed for the light atoms (C, N, O, S, H) in conjunction with the LAN2DZ and LANL2 pseudopotentials and basis set for the heavy species (Pb, I, Br).
The Perdew–Burke-Ernzerhof~\cite{perd+96prl} approximation for the exchange-correlation potential is adopted at this step.
Geometries are relaxed until all interatomic forces are below 4$\times$10$^{-4}$~Ha/bohr.
The following, tabulated values of dielectric functions of the considered solvent are employed within the PCM: $\epsilon_{DMSO}$=46.7, $\epsilon_{DMF}$=38.2, $\epsilon_{ACN}$=36.6, $\epsilon_{NMP}$=32.5, $\epsilon_{GBL}$=41.7, $\epsilon_{PC}$=64.0. 
The values for DMSO, DMF, and ACN are defined by default in Gaussian 16.
The remaining references are taken from the literature~\cite{wohlf15}.
In the subsequent TDDFT calculations, the basis set for the light atoms is tighten to cc-pVTZ. 
The hybrid functional CAM-B3LYP~\cite{yana+04cpl} is employed at this step to approximate the exchange-correlation kernel.

\subsection{Systems}
We construct the considered 18 systems assuming an axial geometry for the PbX$_2$ backbone (see SI, Fig.~S1), which is known to be more stable than the equatorial arrangement~\cite{radi+19acsaem,schi+21pssb}.
Four identical solvent molecules are bound to the Pb atom such that the resulting system is overall in a charge-neutral state. 
The solvent molecules considered in this work can be sorted according to their electron-donating ability~\cite{hami+17acsel}, as shown in Fig.~\ref{fig:solvents}c). 
ACN, which is the only solvent bound to Pb via a N atom, has the lowest electron-donating ability among the chosen ones.
Conversely, DMSO, containing a sulphur atom that connects the oxygen bonding the molecule to the PbX$_2$ backbone to two methyl groups, has the highest tendency to donate electrons.

The resulting complexes are all stable, as discussed in detail in Ref.~\cite{schi+21pssb}, where the same level of theory adopted in this work was employed. 
DMSO as a solvent mostly stabilizes the complexes, leading to formation energies ranging from -0.6~eV in PbCl$_2$ to -0.8~eV in PbI$_2$.
On the other hand, ACN leads to the least stable structures with binding energies on the order of -0.2~eV -- -0.3~eV regardless of the halogen species.
In general, the presence of heavier halogen atoms increases stability.
For further details on energetic and structural properties of these compounds, we refer interested readers to Ref.~\cite{schi+21pssb}.

%%%%%%%%%%%%%%%%%%%%%%%%%%%%%%%%%%%%%%%%
\section{Results}

\subsection{Electronic properties}
\label{sec:electronic}

\begin{figure*}[h!]
	\includegraphics[width=\textwidth]{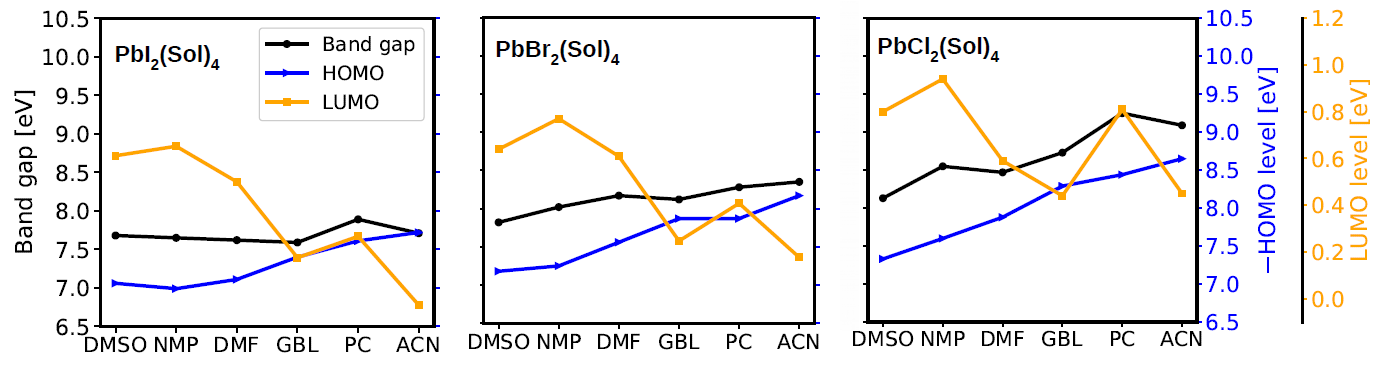}
	\caption{Energies of -HOMO (blue triangles) LUMO (yellow square), as well as their absolute differences (band gaps, black circles) of the compounds considered in this work.}
	\label{fig:gaps}
\end{figure*}

We start the analysis of the electronic properties of the considered compounds by examining the energies of their highest occupied molecular orbital (HOMO) and of the lowest unoccupied molecular orbital (LUMO). 
At a glance, the results reported in Fig.~\ref{fig:gaps} are consistent with chemical intuition: the negative of the HOMO energy increases at decreasing size of the halogen atom while for the LUMO energies the opposite is generally true.
These behaviors can be understood considering that the interaction of the lead-halide cores with electron-withdrawing solvent molecules leads to a stabilization of the HOMO energy, which becomes more negative thereby increasing its absolute value, and to a milder decrease of the LUMO energy (notice the different scales for the HOMO and LUMO energies in Fig.~\ref{fig:gaps}).
An exception in this regard is given by the compounds containing PC, where the LUMO energy is always larger than expected from the trend.

The differences between HOMO and LUMO energies, plotted in Fig.~\ref{fig:gaps} and labeled therein as ``band gaps'' for brevity, are obviously consistent with the picture described above: Larger values are obtained in the compounds with lighter halogen species and with solvent molecules with enhanced electron-withdrawing ability.
Nonetheless, it is worth analyzing these results in more detail, as their rationale is most relevant for the following discussion about optical properties.
The variation of the band gaps with respect to the solvent molecules is much more pronounced in the PbCl$_2$-based compounds than in the others (see Fig.~\ref{fig:gaps}). 
This finding can be understood considering the backbone size. 
Solvent molecules chemically bound to PbCl$_2$ increase significantly the number of electrons in the system. 
On the contrary, the large atomic size of iodine is such that the PbI$_2$ core alone contains 188 electrons.  
In this case, the contribution of the solvent molecules to the electronic cloud represents a small perturbation compared to the one of the inorganic backbone.
Another relevant aspect is related to the influence of the electron-withdrawing solvents to the HOMO-LUMO gaps.
In the presence of DMSO, which is the most electron-donating solvent among the considered ones, the HOMO-LUMO gap varies only within a few hundreds meV depending on the halogen species in the backbone. 
Conversely, the influence of the halogen species is much more pronounced with PC and ACN, the most electron-withdrawing solvents in this study.
This is consistent with their influence on the frontier orbital energies discussed above.
Based on these considerations, we can summarize the driving effects determining the size of the HOMO-LUMO gaps (Fig.~\ref{fig:gaps}) as follows: (i) The increasing size of the halogen atom in backbone leads to a band-gap decrease; (ii) For a given PbX$_2$ backbone, the presence of an electron-withdrawing solvent enhances the band-gap size.
Both effects can be related to the trends obtained for the HOMO, while the behavior of the LUMO is less straightforward.

It is worth underlining that the values of the HOMO and LUMO energies reported in Fig.~\ref{fig:gaps} should not be interpreted as excitation energies. 
In fact, in the context of DFT+PCM calculations, Koopman's theorem is not expected to hold.
The reason for this is the inability of the PCM to correctly capture the electronic reorganization occurring in the system upon ionization (see Ref.~\cite{krum+21pccp} for an extended discussion on this topic in the context of organic semiconductors). 
For this reason, the values of the -HOMO and LUMO energies plotted in Fig.~\ref{fig:gaps} are quantitatively different from the ionization potentials (IP) and the electron affinities (EA) computed for reference within the $\Delta$SCF method, as the difference between the total energies of the neutral species and its cation (IP) or anion (EA) and reported in the SI, Fig.~S2.
Likewise, the HOMO-LUMO gaps differ substantially from the fundamental gaps calculated by subtracting for each system the corresponding EA from the IP (compare Fig.~\ref{fig:gaps} and Fig.~S2).
This given, the qualitative agreement of the trends reported for the two sets of data with respect to the halide species and the solvent molecules confirms that the single-particle energies and their differences can be reliably used in the analysis of optical transitions  (see Section~\ref{ssec:opt}).
As a final remark, we notice that optical excitations computed from the TDDFT+PCM formalism do not suffer from the methodological issues discussed above for the single-particle energies.
The reason behind it is that optical transitions leave the system in a charge-neutral configuration.
To correctly describe this process, the non-equilibrium flavor of PCM~\cite{camm-toma95ijqc} is adopted in the TDDFT calculations. 

\begin{figure*}[h!]
	\centering
	\includegraphics[width=\textwidth]{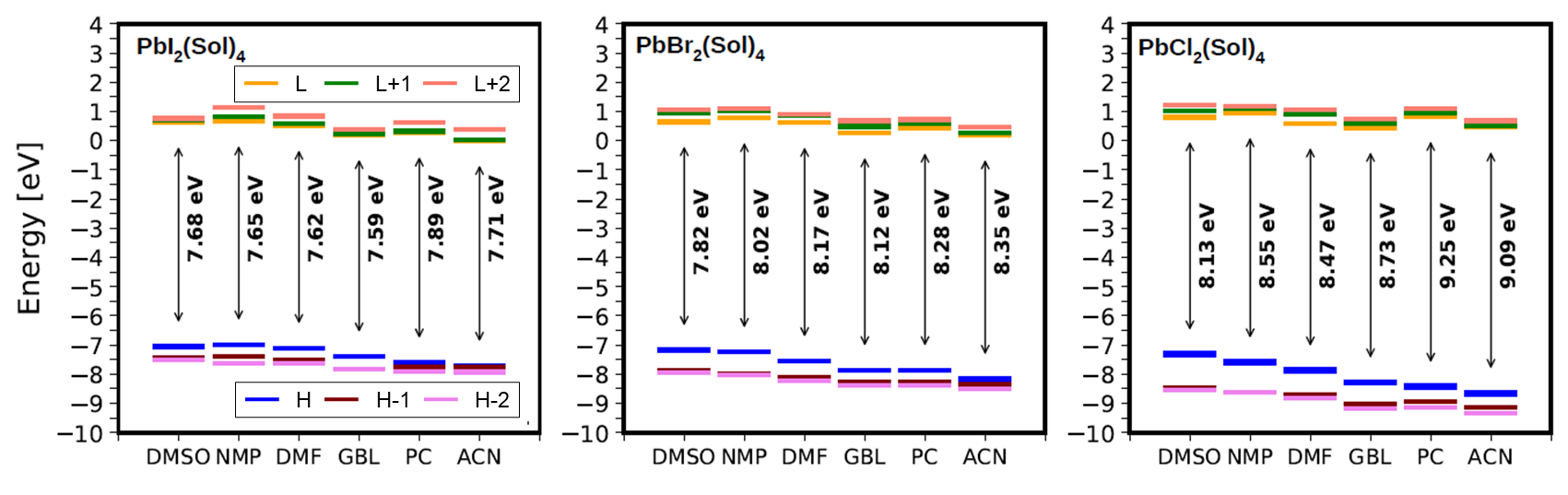}
	\caption{Energy levels close to the frontier of the considered compounds. The values of the band gap are indicated.}
	\label{fig:levels}
\end{figure*}

Before moving to the next part of our study, it is instructive to further extend our analysis to the energy levels in the close vicinity of the HOMO and the LUMO, plotted in Fig.~\ref{fig:levels}.
In most of the compounds considered in this work, the HOMO is energetically separated from the HOMO-1 by a few hundreds of meV, while the latter tends to be closer to the HOMO-2, with energy differences on the order of 100~meV.
This behavior becomes more pronounced with decreasing size of the halogen atoms in the inorganic core.
In all examined systems with PbCl$_2$ backbone, the HOMO is higher in energy than the HOMO-1 by at least 1~eV.
On the other hand, in both PbI$_2$(ACN)$_4$, and PbBr$_2$(ACN)$_4$, as well as in PbI$_2$(PC)$_4$, the HOMO is energetically very close to both the HOMO-1 and the HOMO-2, with the energies of these states spanning a window of less than 400~meV.
In the unoccupied region, the situation is more diversified, although, in general, the LUMO tends to be very close to the LUMO+1 and, in several cases, also to the LUMO+2, with these three states being separated from each other by a few hundreds of meV. 
Exceptions in this regard are given by PbI$_2$(NMP)$_4$ and PbCl$_2$(DMF)$_4$, where the LUMO+1 is above 1~eV in the absolute energy scale reported in Fig.~\ref{fig:levels}, while the LUMO is at approximately 0.6~eV.
All in all, from this analysis, it is evident that electron-withdrawing solvent molecules such as GBL, PC, and ACN tend to lower the HOMO energy, as a signature of electronic stabilization of the compounds induced by their interactions with the backbone.
Conversely, electron-withdrawing solvents, such as DMSO, NMP, and DMF, lead to the opposite effect.
These findings confirm the known trends for LHP solution precursors based on the coordination ability of the various solvents~\cite{hami+17acsel,radi+19acsaem}.

\begin{figure}[h!]
	\centering
	\includegraphics[width=0.9\textwidth]{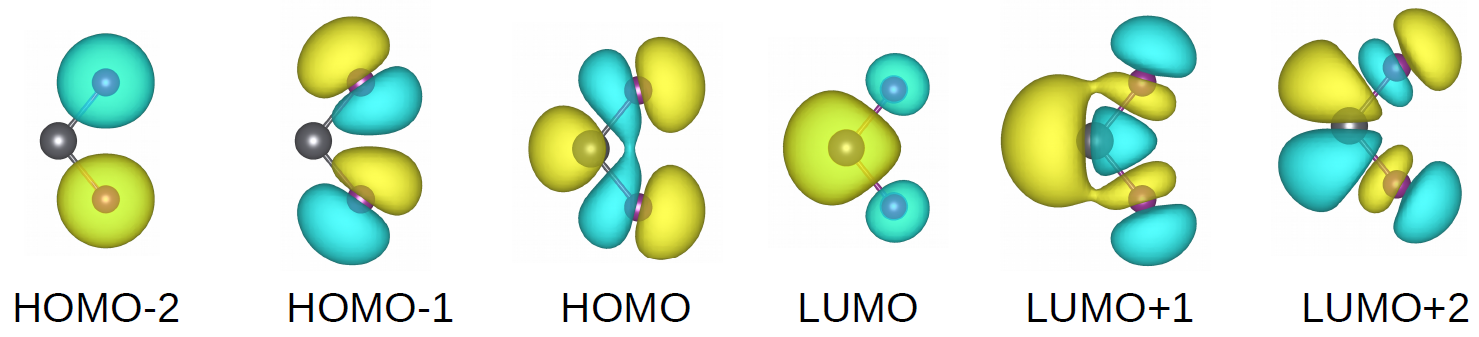} 
	\caption{Isosurfaces of the molecular orbitals computed for PbI$_2$ in an implicit solvent cavity.}
\label{fig:mos}
\end{figure}

The spatial distributions of the molecular orbitals associated with the levels displayed in Fig.~\ref{fig:levels} preserve the characteristics of the corresponding wave-functions in the lead-halide backbone (see in Fig.~\ref{fig:mos} the orbitals of PbI$_2$, which are visually identical to those of PbBr$_2$ and PbCl$_2$, not shown). 
Specifically, the HOMO is formed by a combination of a $p$-state in the halogen atom and of the $6s$ state in Pb, while the HOMO-1 and the HOMO-2 include only contributions from halogen $p$-orbitals (no charge on Pb).
In the unoccupied region, the first three molecular orbitals are dominated by the hybridization between Pb and halogen $p$-states.
In the solvated compounds, the chemical bond between the PbX$_2$ core and the solvent molecules~\cite{schi+21pssb} promotes charge delocalization to the latter (see Figs.~S3 -- S8 in the SI). 
However, these electronic interactions act mainly on the energy of the states, as detailed in Fig.~\ref{fig:levels}, rather than on the nature of the orbitals, as anticipated above.
In terms of wave-function distribution, the most prominent effects of solute-solvent hybridization consist in a slight charge delocalization the across the Pb-O bonds (or across the Pb-N bonds in ACN). 
Interestingly, these features are generally preserved in all systems, although the actual distribution of the orbitals depends on the details of solute-solvent hybridization and also on the size of the halide species
(see Figs.~S3 -- S8).
In general, these interactions tend to be more pronounced in the unoccupied levels and are more prominent in PbCl$_2$-based compounds due to the smaller size of the Cl atom compared to the other considered halide species.
With only 17 electrons distributed up to the 3$s$ and 3$p$ shells, the chemical environment provided by Cl to the Pb atom is not very different compared to the solvent molecules.
As a result, the occupied electronic levels below the HOMO as well as the unoccupied ones become energetically close to each other, as discussed above with reference to Fig.~\ref{fig:levels}, and the corresponding orbital distribution is more homogeneously spread over the entire compound.

%%%%%%%%%%%%%%%%%%%%%%%%%%%%%%%%%%%%%%%%
\subsection{Optical properties}
\label{ssec:opt}

\begin{figure*}[h!]
	\centering
	\includegraphics[width=\textwidth]{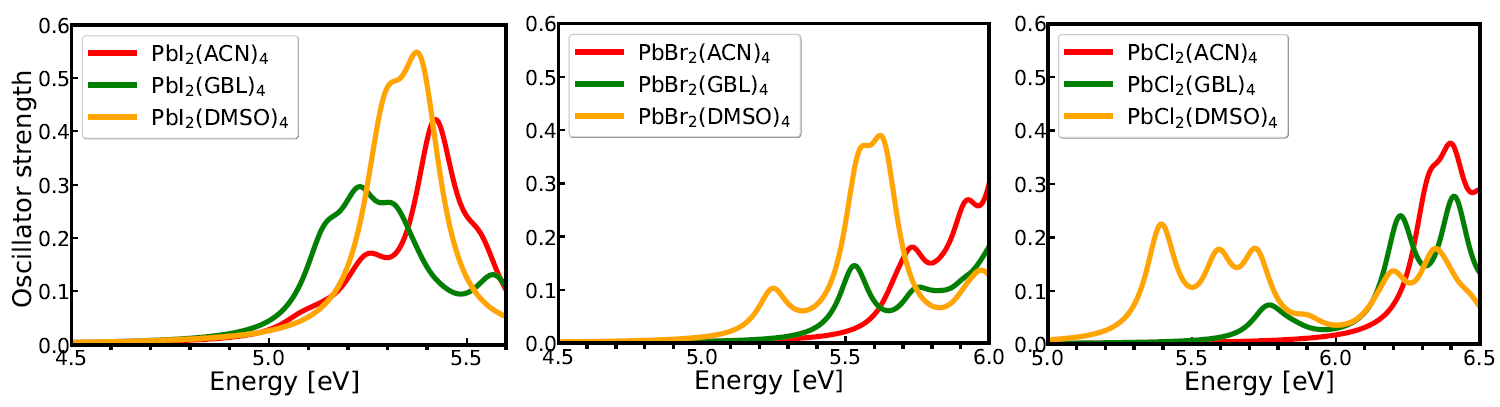} 
	\caption{Optical absorption spectra of PbX$_2$DMSO$_4$, PbX$_2$GBL$_4$, and PbX$_2$ACN$_4$ with X~=~I (left), X~=~Br (middle), X~=~Cl (right). A Lorentzian broadening of 125 meV is applied to all spectra.}
\label{fig:spectra}
\end{figure*}

We start the analysis of the optical properties from the computed absorption spectra. 
We focus specifically on the lead-halide compounds interacting with the most electron-donating solvent (DMSO), the most electron-withdrawing one (ACN), as well as GBL, which sets itself between the former two (see Fig.~\ref{fig:solvents}a).
The optical spectra of these systems are plotted in Fig.~\ref{fig:spectra}, while those of the remaining compounds are reported in the SI, Fig.~S9.
From the trends depicted in Figs.~\ref{fig:gaps} and \ref{fig:levels}, it is reasonable to expect that the absorption onset increases in energy with decreasing size of the halogen atoms, while it decreases with increasing electron-donating ability of the solvent molecules.
The results shown in Fig.~\ref{fig:spectra} are only partly consistent with this intuition.

The spectra of the compounds with PbBr$_2$ and PbI$_2$ backbones exhibit the first peaks at higher energies compared to their PbCl$_2$-based counterparts (notice the different energy scales in the three panels of Fig.~\ref{fig:spectra}).
The increasing electron-donating ability of the solvents is reflected in the red-shift of the absorption onsets, as expected based on the results in Figs.~\ref{fig:gaps} and \ref{fig:levels}.
However, in the spectra of the PbX$_2$(DMSO)$_4$ compounds, the absorption onset undergoes a red-shift of $\sim$200~meV upon decreasing size of the halogen species, consistent with the results obtained for the electronic gaps (Fig.~\ref{fig:gaps}). 
In the spectra of the PbI$_2$-based systems (Fig.~\ref{fig:spectra}, left panel), the lowest-energy onset corresponding to the compound in GBL is in line with the trends of the HOMO-LUMO gaps (Figs.~\ref{fig:gaps} and \ref{fig:levels}), although in absolute terms the difference between the absorption onset energies is larger compared to the band gap energies. 
This discrepancy can be explained recalling the issues with the quantitative interpretation of the single-particle energies calculated within the DFT+PCM formalism (see end of Section~\ref{sec:electronic} and Ref.~\cite{krum+21pccp}).
The qualitative agreement between the two sets of data is therefore satisfactory. 

\begin{figure*}[h!]
	\centering
	\includegraphics[width=0.98\textwidth]{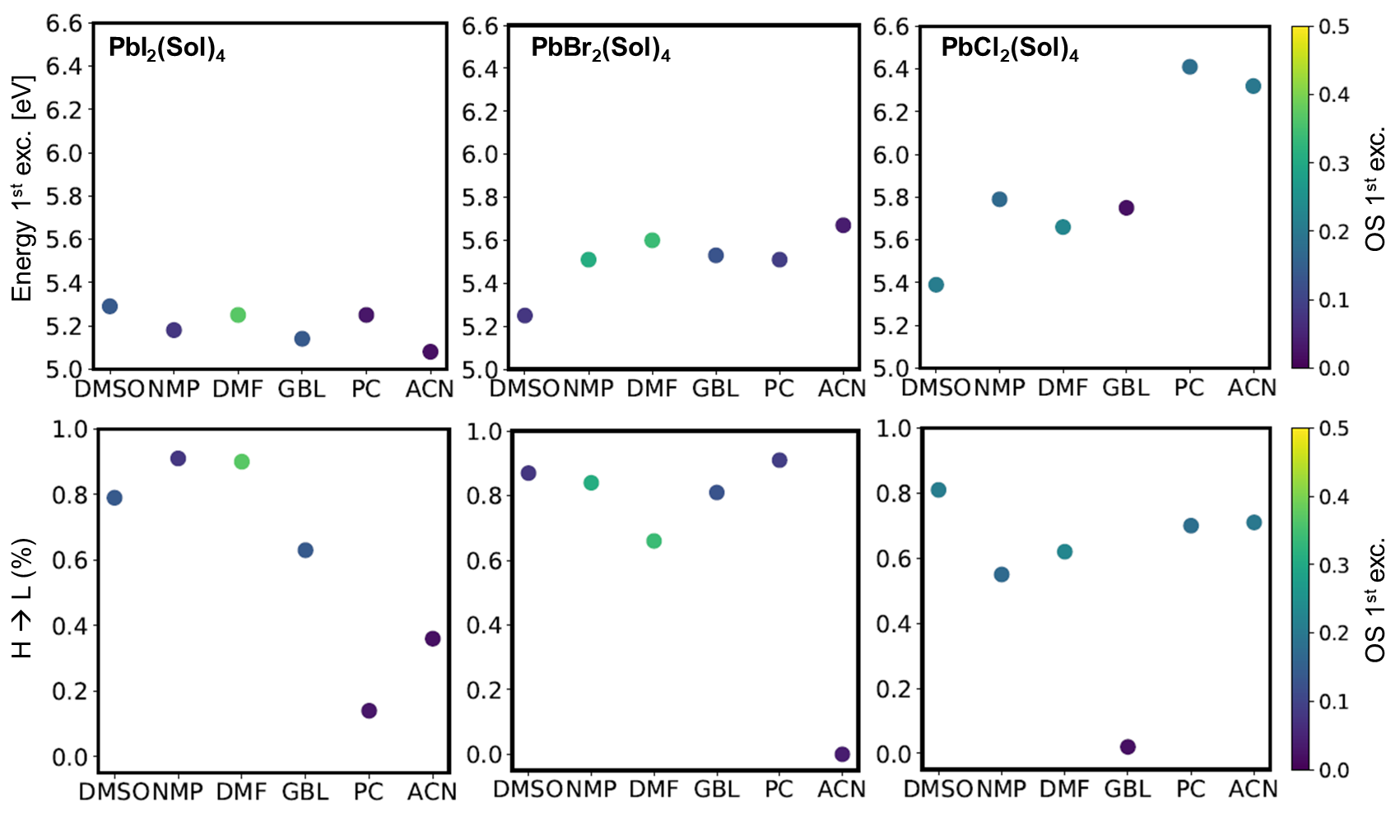}
		\caption{Analysis of the first excitation in all the considered systems, from left to right, PbI$_2$(Sol)$_4$ , PbBr$_2$ (Sol)$_4$, and PbCl$_2$(Sol)$_4$. Top: Excitation energy; Bottom: Fractional amount of the HOMO $\rightarrow$ LUMO transitions contributing to the first excitation. The color bar indicates the oscillator strength (OS). }
	\label{fig:first}
\end{figure*}

To gain further insight into the optical properties, we analyze more closely their lowest-energy excitations.
A visual overview is provided in Fig.~\ref{fig:first}, top panel, for all systems investigated in this work, while all details are reported in the SI, Tables~S1 -- S18.
In the PbI$_2$-based compounds, the energies of first excitations are comprised within a window of less than 300~meV, while in the systems with PbBr$_2$, backbone they cover a twice as large energy range; finally, in the chlorinated materials, the energies of the first excited states vary within approximately 1~eV. 
This trend is consistent with the energy distribution of the HOMO-LUMO gaps (Fig.~\ref{fig:gaps}) and can be explained analogously (see Section~\ref{sec:electronic}): Solvent molecules bound to the PbCl$_2$-based compounds increase the number of electrons more dramatically than with the PbI$_2$ cores, thereby affecting more prominently the distribution of the single-particle states and hence the transition energies between them.
Yet, excitation energies are \textit{per se} insufficient to rationalize optical spectra: the oscillator strengths (OS) of the excited states and their composition in terms of single-particle transitions provide valuable information too.

Focusing on the PbI$_2$(Sol)$_4$ compounds first (Fig.~\ref{fig:first}, top left panel), we notice that the system bound to DMF exhibits the most intense first excitation; in the case of PbI$_2$(DMSO)$_4$ and PbI$_2$(GBL)$_4$, the oscillator strength of the lowest-energy peak is lower than in PbI$_2$(DMF)$_4$ but still non-negligible; finally, in PbI$_2$(NMP)$_4$, PbI$_2$(PC)$_4$, and especially PbI$_2$(ACN)$_4$, the first excitation can be considered almost dark, having an OS below 0.1 (see SI for the corresponding values).
The reason for the different OS exhibited by the first excitation of these compounds can be traced back to their composition in terms of single-particle transitions.
By inspecting once again the molecular orbitals shown in Fig.~\ref{fig:mos}, it is clear that transitions from the HOMO to the LUMO are optically allowed, due to the $s$- and $p$-character of the contributions localized on the Pb atom in the respective wave-function distributions; this characteristic is preserved also in the presence of the solvent molecules (see Figs.~S3 -- S8 in the SI).
For this reason, we analyze the relative weight of the HOMO$\rightarrow$LUMO transition to the lowest-energy excitation in each considered system.

From the corresponding plots, shown on the bottom panel of Fig.~\ref{fig:first}, the off-trend behavior exhibted by PbI$_2$(ACN)$_4$, PbI$_2$(PC)$_4$, PbBr$_2$(ACN)$_4$, and PbCl$_2$(GBL)$_4$ is clearly noticeable: in all these systems, the first excitation receives a low or negligible contribution from the HOMO$\rightarrow$LUMO transition, and the OS is well below 0.1 (see Tables~S1, S2, S7, S15).
This behavior can be understood considering the electronic structure of these compounds.
Going back to Fig.~\ref{fig:levels}, we notice that in PbI$_2$(ACN)$_4$, in PbI$_2$(PC)$_4$, and in PbBr$_2$(ACN)$_4$, both frontier orbitals are very close to the neighboring energy levels, while for PbCl$_2$(GBL)$_4$ this is true for the LUMO only.
As a result, several single-particle transitions contribute to the first excited state, with a predominance of HOMO-1$\rightarrow$LUMO in PbI$_2$(ACN)$_4$, PbI$_2$(PC)$_4$ and in PbBr$_2$(ACN)$_4$.
In the case of PbCl$_2$(GBL)$_4$, the situation is even more complex, as more than five transitions all with weights below 15\% contribute to the first excitation.
Based on the distribution of the HOMO-1 and of the LUMO in the lead halide backbone (see Fig.~\ref{fig:mos}) their transition is expected to be optically inactive (see also Tables~S19 -- S21 in the SI).
The interaction with the solvent molecules promotes charge delocalization onto the latter and breaks the symmetry of the inorganic core: as a result, the OS in these systems is almost one order of magnitude larger than in the isolated PbX$_2$ units.
Nonetheless, it is clear from Fig.~\ref{fig:spectra} (and Fig.~S2 for PbI$_2$(PC)$_4$) that the first excitation does not contribute substantially to the optical absorption of these systems. 
As a side note, we mention that the same behavior for PbI$_2$(PC)$_4$ is reproduced by many-body perturbation theory (MBPT) calculations \textit{in vacuo}  (see Table~S22 in the SI).
In the case of PbI$_2$(ACN)$_4$, MBPT predicts the first excitation to stem almost entirely from the HOMO$\rightarrow$LUMO transition.
However, in this compound, the quasi-particle correction to the orbital energies computed from $G_0W_0$ switches the order for the HOMO and HOMO-1 (see Fig.~S11), such that, effectively, the first excitation in both PbI$_2$(PC)$_4$ and PbI$_2$(ACN)$_4$ is dominated by the same orbital transition.

Inspecting now the other results plotted in Fig.~\ref{fig:first}, bottom panel, we notice that in all compounds except those discussed above, the weight of the HOMO$\rightarrow$LUMO transition to the first excitation is above 50\% and, correspondingly, the OS is close to or above 0.1 (see SI for details). 
Indeed, there is a clear correlation between the OS and the contribution of the transition between the frontier orbitals.
The most remarkable example is given by the compounds interacting with DMF: in all these three systems, the first excitation has an oscillator strength above 0.25 and a relative HOMO$\rightarrow$LUMO weight larger than 60\%.

From the analysis of the optical properties for the PbX$_2$ compounds computed in an implicit solvent cavity but without explicit solvent molecules, we find that the behaviors rationalized above stem directly from the electronic and optical properties of the lead halide backbones.
As detailed in the SI, on Tables~S19 -- S21, also in these systems the lowest-energy excitation is bright, as visible in the optical spectra (Fig.~S10), and it is almost entirely given by the HOMO$\rightarrow$LUMO transition.
However, in the absence of chemically bound solvent molecules, the associated OS is only of the order of 10$^{-2}$, suggesting that the presence of the latter actually enhances the optical activity of this transition.  
The most intense excitation is associated with the HOMO$\rightarrow$LUMO+1, which gives rise to the strong peak, which is found at least 1.5~eV above the onset depending on the halogen species. 
The relatively large OS associated with the HOMO$\rightarrow$LUMO+1 transition is present also in the explicitly solvated compounds.
In most cases, it also appears at higher energies compared to the first bright excitation, such that it does enter the range plotted in Figs.\ref{fig:spectra} and S9 (see Tables~S1 -- S18 for details). 
However, due to the higher density of excited states in the PbX$_2$(Sol)$_4$ compounds compared to the PbX$_2$ ones, the energy differences between the excitations dominated by the HOMO$\rightarrow$LUMO and the HOMO$\rightarrow$LUMO+1 transitions are within 1~eV.
There are nonetheless some exceptions to this behavior, with the most significant one concerning the systems interacting with DMSO. 
Regardless of the halogen atom in the backbone, in these compounds, the second excitation is dominated by the HOMO$\rightarrow$LUMO+1 contribution and it is therefore as intense as the first one in the spectra of PbI$_2$(DMSO)$_4$ and PbCl$_2$(DMSO)$_4$, and even more intense in the spectrum of PbBr$_2$(DMSO)$_4$ (see Fig.~\ref{fig:spectra}). 

Based on this analysis, the correlation between the oscillator strength of the first excitation in the PbX$_2$(Sol)$_4$ systems and the relative contribution to it from the HOMO$\rightarrow$LUMO transition (see Fig.~\ref{fig:first}, bottom panels) can be understood as follows.
The presence of solvent molecules chemically bound to the PbX$_2$ backbone does not affect substantially the character of the frontier orbitals reported in Fig.~\ref{fig:mos} (see Figs.~S3 -- S8 in the SI). 
Hence, the HOMO$\rightarrow$LUMO transition, which is optically allowed in PbX$_2$ (see Table~S19 in the SI) due to the symmetry and parity of their orbitals (Fig.~\ref{fig:mos}), remains active also in the solvated complexes, PbX$_2$(Sol)$_4$. 
On the contrary, both the HOMO-1$\rightarrow$LUMO and the HOMO$\rightarrow$LUMO+1 transitions have negligible oscillator strength in the PbX$_2$ units, again owing to the character of the involved orbitals (Fig.~\ref{fig:mos}). 
Hence, when the lowest-energy excitation in PbX$_2$(Sol)$_4$ is dominated by a large fraction of the HOMO$\rightarrow$LUMO transition, its oscillator strength is large.
When this does not happen and an optically inactive transition is dominant, as in the case of PbI$_2$(ACN)$_4$, PbI$_2$(PC)$_4$, PbBr$_2$(ACN)$_4$, and PbCl$_2$(GBL)$_4$ (see Tables~S1, S2, S7, and S15 in the SI), the optical spectrum exhibits dark excitations at lowest-energy (Fig.~\ref{fig:first}, bottom panels).
We emphasize that the energetic proximity between the HOMO and/or the LUMO and the neighboring single-particle states in these systems (see Fig.~\ref{fig:levels}) promotes this behavior.

For further comparison, we analyze the optical spectra computed for the PbI$_2$-based compounds in the framework of MBPT~\cite{weig+02,brun12jcp,brun+16cpc} without the inclusion of implicit solvation effects.
The same approach was successfully applied in Ref.~\cite{vale+21jpcl} in a combined computational/experimental study.
The trends provided by these results (see SI, Fig.~S9, S10, and Table~S19) are consistent with those obtained from TDDFT+PCM: larger electronic and optical gaps are associated with compounds with electron-donating groups.
We notice, however, that the absolute values of the excitation energies computed from MBPT are systematically lower compared to those in Fig.~\ref{fig:spectra} (see Fig.~S12).
This can be explained considering that the quasi-particle correction included via the $G_0W_0$ approximation on top of the DFT results (CAM-B3LYP functional) increases the band-gap size.
On top of this, the electron-hole interaction explicitly accounted for in the solution of the Bethe-Salpeter equation red-shifts the absorption onset by more than 3.5~eV in each system.
This value, which is interpreted as the exciton binding energy, is of the typical order of magnitude of small molecules~\cite{hiro+15prb,cocc-drax15prb}.

Optical absorption spectra measured for lead iodide perovskite precursors in different solvents are in overall agreement with our findings~\cite{shar+20ma}.
It is important to recall that in actual samples, different species coexist and contribute to the detected optical absorption spectra~\cite{rahi+16chpch,radi+19acsaem,shar+20ma,vale+21jpcl}.
Nonetheless, we can comment qualitatively on the fingerprints associated to the neutral PbI$_2$(Sol)$_4$ species therein.
In the measurements reported in Ref.~\cite{shar+20ma}, the peak energies assigned to the lowest-energy transitions in PbI$_2$(GBL)$_4$, PbI$_2$(DMSO)$_4$, PbI$_2$(NMP)$_4$, and PbI$_2$(DMF)$_4$ are almost at the same energy, which is not in contrast with our findings, considering the inhomogeneous broadening in the experimental spectra.
Interestingly, in the same work, the peak associated to the PbI$_2$(DMSO)$_4$ species is much more intense than its counterpart for PbI$_2$(GBL)$_4$~\cite{shar+20ma}, in qualitative agreement with our results shown in Fig.~\ref{fig:spectra}, left panel.
Even in Ref.~\cite{radi+19acsaem}, a lower absorption onset of PbI$_2$(GBL)$_4$ compared to PbI$_2$(DMSO)$_4$ is reported for measurements of these compounds in the presence of the methylammonium cation.
The systematic blue-shift of our computed spectra with respect to experimental references is to be mainly ascribed to the absence in our calculations of the non-negligible contributions of spin-orbit coupling~\cite{borg+19ctc} and to the absence of additional intermolecular interactions beyond those between the PbX$_2$ backbones and the nearest-neighboring solvent molecules.
However, as demonstrated in Ref.~\cite{vale+21jpcl}, a quantitative misalignment with experimental absorption energies does not affect the insight provided by theory in the interpretation of the measured spectra.

%%%%%%%%%%%%%%%%%%%%%%%%%%%%%%%%

\section{Summary, Conclusions, and Outlook}
In summary, we have presented a first-principles study of the electronic and optical properties of a representative set of charge-neutral LHP solution precursors with chemical formula PbX$_2$(Sol)$_4$ (Sol = DMSO, PC, NMP, DMF, GBL, and ACN). 
Our analysis reveals the correlations between the electron-donating ability of the solvent molecules and the decreasing size of the band gap of the solutes, due to a concomitant decrease (increase) of the HOMO (LUMO) energy.
Notably, this trend is highly sensitive of the halogen atoms in the backbone: in the presence of iodine, band gaps span an energy window of less than 0.5~eV; when the solute contains Br and especially Cl, the band-gap values of the investigated systems vary within a range of approximately 1~eV.
Orbital hybridization between solute species and explicit solvent molecules occur mainly in the unoccupied region and is more prominent in the PbCl$_2$-based compounds due to the smaller number of electrons in chlorine atoms compared to the other considered halogen species. 
However, these effects do not alter the nature of the molecular orbitals, which preserve their character as in the lead halide backbone.
The inclusion of heavier halogen atoms in the inorganic core leads to a red-shift of the absorption onset, in accordance with the aforementioned trends of the HOMO-LUMO gaps.
Except for a few compounds with electron-withdrawing solvents such as PbI$_2$(PC)$_4$, PbI$_2$(ACN)$_4$, PbBr$_2$(ACN)$_4$, and PbCl$_2$(GBL)$_4$, in all systems, the lowest-energy excitation is bright and corresponds to the optically active HOMO$\rightarrow$LUMO transition.
The correlation between the weight of this contribution to the first excited state and the oscillator strength of the latter is key to understand the optical absorption of these compounds.

In conclusion, with this study, we have clarified the fundamental relations between absorption spectra and electronic structure of LHP solution precursors, and highlighted how the halogen atoms in the inorganic core as well as the solvent molecules can critically impact on the optical response of these compounds. 
Our findings are useful to identify and interpret the fingerprints of the neutral species PbX$_2$(Sol)$_4$ in optical absorption measurements, where the contributions of these compounds are entangled with those of lead halide species with different coordination numbers that coexist in the samples.
The disclosed non-trivial effects of the solvent molecules on the single-particle levels close to the frontier suggest that a simplistic interpretation of the optical absorption based solely on the molecular-orbital energies is insufficient to interpret the spectra of LHP solution precursors.
While the optically-active HOMO$\rightarrow$LUMO transition is often dominating the first excitation, in the cases when this is not true the insight provided by our study can be of great help.  

The theoretical framework adopted herein and the presented analysis can be successfully extended also to other lead halide species that are detected in the samples, as well as to LHP intermediate structures identified at subsequent stages of formation.  
As such, the outcomes of this study crucially contribute to the understanding of solute-solvent interactions in electronic and optical properties of LHP solution precursors.
Furthermore, they provide relevant indications to assess the role of spurious solvent molecules that are still embedded in LHP intermediates during their evolution to crystalline thin films~\cite{guo+15jacs,cao+16jacs,petr+17jpcc}.
Dedicated study will be necessary to further disclosed the electronic structure of such systems and their interaction with (visible) light.
%%%%%%%%%%%%%%%%%%%%%%%%%%%%

\section*{Data Availability Statement}
The data reported in this work are publicly available open access in the Zenodo repository at the following link: https://doi.org/10.5281/zenodo.5416968. 

%%%%%%%%%%%%%%%%%%%%%%%%%%%%%%%%%
%
%\section*{Conflicts of interest}
%There are no conflicts to declare.
%%%%%%%%%%%%%%%%%%%%%%%%%%%%%%%%%%%%%%%%%

\section*{Acknowledgements}
We are grateful to Jannis Krumland, Michele Guerrini, Oleksandra Shargaieva, and Eva Unger for fruitful discussions.
This work was supported by the German Research Foundation (DFG), Priority Programm SPP 2196 - Project number 424394788, by the German Federal Ministry of Education and Research (Professorinnenprogramm III), and from the State of Lower Saxony (Professorinnen für Niedersachsen).
Calculations were performed on the HPC cluster CARL at the University of Oldenburg, funded by the DFG (Project No. INST 184/157-1 FUGG) and by the Ministry of Science and Culture of the State of Lower Saxony.

%%%END OF MAIN TEXT%%%

%\section*{References}
%\bibliography{Ref}% Produces the bibliography via BibTeX.


\end{document}